This is the beginning of the LaTeX file.

\documentstyle[12pt]{article}
\catcode`\@=11 \@addtoreset{equation}{section}\catcode`\@=12
\begin{document}
\thispagestyle{empty}

\begin{center}

               RUSSIAN GRAVITATIONAL ASSOCIATION\\
               CENTER FOR SURFACE AND VACUUM RESEARCH\\
               DEPARTMENT OF FUNDAMENTAL INTERACTIONS AND METROLOGY\\
\end{center}
\vskip 4ex
\begin{flushright}                              RGA-CSVR-008/94\\
                                                gr-qc/9406051
\end{flushright}
\vskip 45mm

\begin{center}
{\bf
On description of spatial topologies in quantum gravity}

\vskip 5mm
{\bf A. A. Kirillov}\\

\vskip 5mm
{\em Institute for Applied Mathematics and Cybernetics\\
10 Uljanova str., Nizhny Novgorod, 603005, Russia}\\
     e-mail: Kirillovl@focus.nnov.su\\
\vskip 60mm

           Moscow 1994
\end{center}
\pagebreak

\setcounter{page}{1}
A new approach is suggested which allows to describe phenomenologically
arbitrary topologies of the Universe. It consists in a
generalizaton of third quantization. This quantization is carried out for the
case of asymptotic closeness to a cosmological singularity. It is also
pointed out that suggested approach leads to a modification of the ordinary
quantum field theory. In order to show this modification we consider the
example of a free massless scalar field.\\

\section{Introduction}
It is widely accepted that quantum fluctuations of metrics at small-scale
distances can change spatial topology of the Universe [1],[2].
Effects connected with topology changing have been considered already in
Refs.[3-6].
Nevertheless, an adequate mathematical scheme for description
of processes of such a kind is absent yet. In this paper we suggest an approach
which, as we hope, gives a possibility
for, at least, a phenomenological description of arbitrary spatial topologies.
To this end we use a generalization of third quantization.

Third quantization has been already used in quantum cosmology for description
of "wormholes" and "baby universes" [3-6] (which lead as it was shown to the
loss
of quantum coherence) as well as for description of "spontaneous quantum
creation of a universe from nothing" [7],[8] proposed earlier in Ref.[9].
Note however that in the all considered cases a number of small closed
universes (which branch off the our large Universe)
is a variable quantity but it turns out that the topology of each closed
universe is fixed.
To describe different possible topologies one should modify the
procedure of the third quantization. The simplest way to do this is to make it
local.
Such a possibility follows from the fact that the Wheeler-DeWitt equation
which governs the evolution of a wave function of the Universe consists of
an infinite set of the Klein-Gordon type equations (one local Wheeler-DeWitt
equation at each point $x$ of three-space $S$). We note that this fact is in
accordance with another one that in General Relativity time has only a local
meaning.
Therefore, one may attempt to quantize every local Wheeler-DeWitt equation
independently. Such procedure we call the {\it Local Third Quantization} (LTQ).

There is one problem with LTQ procedure in general case. The fact is that all
local Wheeler-DeWitt equations are strongly coupled to each other and so it is
very difficult to carry out this procedure. Nevertheless there is a situation
when connection between different local Wheeler-DeWitt equations disappear,
at least, in main order. It is just the case if one treats a behavior of a
gravitational field close to a cosmological singularity.

It was shown in Refs.[10,11] that a general inhomogeneous gravitational field
at the singularity may be considered as a continuum of uncoupled homogeneous
(of IX type) fields. Indeed, nearby the singularity it is always possible to
choose an "elementary" volume $\Delta V$, in which the gravitational field is
homogeneous in leading order (we do not take into account the presence of
matter for the sake of simplicity and because it does not change properties of
the gravitational
field). In the vicinity of the singularity the horizon size tends to
zero $(l_{h}\rightarrow 0)$ and different "elementary" domains $\Delta V(x)$
of the three-space $S$, do not affect on each other and may be considered
independently (for validity of that the following condition must be fulfilled:
$(\Delta V)^{1/3}\ll l_{h})$. The LTQ procedure consists then of the assumption
that quantization is carried out independently for each elementary spatial
domain $\Delta V(x)$. Furthermore, one may assume that all these domains are
indistinguishable. Localization of third quantization is achieved in the limit
as $\Delta V\rightarrow 0$ only. Notice that in this limit every such
"elementary"
domain contains only one "physical point" of space and, therefore,
under the "elementary" domain we will mean an isolated point of physical
continuum.

The {\it Local Third Quantization} leads to a modification of the ordinary
quantum field theory. Indeed, in the limit $\Delta V\rightarrow 0$ an
"elementary" domain $\Delta V(x)$ contains a finite number of physical
degrees of freedom which coincides with the number of physically arbitrary
functions determining distributions of matter and gravitational field.
At each point of $S$ these degrees of freedom form a set. Thus, under the {\it
Local Third Quantization} one may mean the independent third quantization
of all such sets. In ordinary field theory it is more convenient to use Fourier
transformation for physical degrees of freedom that is expansion in modes.
So in the case of free fields local third quantization consists in third
quantization of the field modes.

\section{A Gravitational Field in the Vicinity of a Cosmological Singularity}

As it was pointed out above the problem of the local third quantization of the
gravitational field at the singularity is reduced to the third
quantization of the homogeneous field. It was shown in Refs.[8] and [11]
(see also Ref.[12]) that at the singularity the quantum states of the
homogeneous field (or in our case of an "elementary" spatial domain
$\Delta V(x)$) may be classified by some quantum number $n (n=0,1...)$.
When third quantization is imposed, the wave function becomes a field
operator and can be expanded in the form
(here for simplicity we assume that $\Psi $ is a real scalar function)
\begin{equation}
\Psi  = \sum C_{n}U_{n}+ C^{+}_{n}U^{*}_{n},
\end{equation}
where $\{U_{n},U^{*}_{n}\}$ is an arbitrary complete orthonormal set
of solutions of the Wheeler-DeWitt equation:
\begin{equation}
(\Delta +V) U_{n} = 0
\end{equation}
here $V$ is a potential, $\Delta ={1\over \sqrt{-G}}\partial
_{A}\sqrt{-G}G^{AB}\partial _{B}$ and $G_{AB}$
is the metric on a minisuperspace $W$ (for more detail see Ref.[11]).
The operators $C_{n}$ and $C^{+}_{n}$ satisfy the standard (anti)commutation
relations
\begin{equation}
[C_{n},C^{+}_{m}]_{\pm }= \delta_{n,m},
\end{equation}
where $\pm $ relates to the two possible statistics of the wave function.

In the case of inhomogeneous field
the Wheeler-DeWitt equation is splited up into the set of the uncoupled
equations of the (2.2)
type, each of which contains the variables describing a gravitational field at
a fixed
point $x$ of the three-manifold $S$:
\begin{equation}
(\Delta(x) +V_{x}) \Psi=0.
\end{equation}
The space $H$ of solutions of the Wheeler-DeWitt equation has the form of the
tensor product
of the spaces $H_{x}$ (written as $H =\prod_{x\in S}H_{x})$ where $H_{x}$is the
space of solutions of Eq.(2.2).
Then one may introduce a set of wave functions $\Psi _{x}$ and secondly
quantize
every local Wheeler-DeWitt equation (2.4) independently.
Thus the operators (2.3) acquire additional
dependence of spatial coordinates. The LTQ procedure consists thus in the
replacement of the relations (2.3) by the following ones
\begin{equation}
[C(x,n),C^{+}(y,m)]_{\pm }=\delta _{n,m}\delta (x-y).
\end{equation}

Using the operator algebra (2.5) one can construct a set of states with an
arbitrary number of domains (with an arbitrary density of points for physical
continuum). In particular,
the vacuum state is determined
by the relations $C(x,n)\mid 0>=0$ (for arbitrary $x\in S$) and, therefore,
this state corresponds to the absence of all points of physical space
and consequently the absence of all field observables.
In other words in this state there is no a physical continuum
as it is. The operator
$N(x,n)=C^{+}(x,n)C(x,n)$ has the ordinary meaning of number of elementary
domains $\Delta V(x)$ given in the quantum state
$U_{n}$ and located at the point $x\in S$.
Summarizing over $n$ one may found the complete number operator of domains
having the coordinate  $x: N(x)=\sum N(x,n)$ which has sense of a operator of
density of physical points).
The operator $\theta (x)=1-N(x)$ may be used then as an indicator of
difference of topology of the Universe from that of $S$.

Considered theory includes conventional quantum gravity as a
particular case. Indeed, let us consider the set of states
$\{\mid A>\}$ which have the form
\begin{equation}
\mid [n(x)]>=\prod^{}_{x\in S}C^{+}(x,n(x))\mid 0>.
\end{equation}
These states describe the case when there is just one elementary domain at
each point $x\in S$ and, therefore, the following conditions are fulfilled:
\begin{equation}\theta (x)\mid A>=0,  as  x\in S
\end{equation}
(i.e. the number of point of physical continuum having the coordinate $x$
coincides with that of the $S$). Obviously that for these states topology of
physical space is the same as
that of the $S$.

In order to illustrate a nontrivial topology of the Universe
one may construct a handle on $S$. In our approach the existence of the handle
is indexed by the fact that quantum states of the gravitational field
$U_{n(x)}$
are triple-valued functions of spatial coordinates (in some region
$K\in S)$. Therefore, the states describing the handle may be taken in
the form
$$
\mid [n(x)];[p(x)],[q(x)]_{K}>=\prod^{}_{y\in
K}C^{+}(y,p(y))C^{+}(y,q(y))\prod^{}_{z\in S}C^{+}(z,n(z))\mid 0>.
$$

It is obvious that due to indistinguishability of domains one may speak about
topology of physical space in a usual sense in quasi-classical limit only.
Indeed, in this limit one can introduce a set of maps such that metric
functions
become single-valued.

Evidently, one of possible applications of LTQ is a
description of effects connected with the "space-time foam"[1], [2].
In particular, it should display itself in the existence of
the so-called vacuum fluctuations connected with the creation and annihilation
of virtual points of physical space. It should be also noted that the numbers
$N(x)$ vary during the evolution [7], [8]. This means that the structure
of the foam is not fixed and must be determined dynamically.
Besides, there is an interesting possibility that at small distances the
spatial continuum has "hollows" (i.e. $N({\bf k})\rightarrow 0$
if ${\bf k}\rightarrow \infty $,
where $N({\bf k})=(2\pi )^{-2/3}\int^{}{}N(x)\exp(-i{\bf k}x)d^{3}x)$. Thus,
in this way, one may attempt to overcome the divergences problem in
conventional
quantum gravity.

\section{On a Modification of the Ordinary Field Theory}

The foamy structure of the spacetime must be reflected in a universal way
on the structure of the conventional field theory. As an example we consider
now a free massless scalar field $\varphi $.

If one writes the Fourier expansion for $\varphi $
\begin{equation}\varphi (x,t)=(2\pi )^{-2/3}\int {d^{3}{\bf k}\over
\sqrt{2k}}\left\{\begin{array}{l}
A({\bf k})e^{i{\bf k}x\hbox{{\it -ikt}}}+A^{+}({\bf k})e{ } ^{-i{\bf
k}x+\hbox{{\it ikt}}}\end{array}\right\}
\end{equation}
(here $k=\left|\begin{array}{c}
{\bf k}\end{array}\right| )$, then the Hamiltonian of the field takes the
form of the sum of independent non-interacting oscillators
\begin{equation}H=\int {k\over 2}\left\{\begin{array}{l}
A({\bf k})A^{+}({\bf k})+A^{+}({\bf k})A({\bf k})\end{array}\right\}d^{3}{\bf
k}.
\end{equation}
Since, as it was mentioned in sec.2, the number of spatial domains $N({\bf k})$
may be a variable quantity so does the number of field oscillators. This fact
may be accounted phenomenologically by introducing of the creation and
annihilation
operators of field oscillators which obey the same (anti)commutation relations
as in (2.5)
\begin{equation}[C({\bf k},n),C^{+}({\bf k}',m)]_{\pm }=\delta _{n,m}\delta
^{3}({\bf k}-{\bf k}')
\end{equation}
where dependence of the operators upon the quantities ${\bf k}$ and $n$ is
connected with the
classification of the states of a separate oscillator (the spectrum of the
oscillator has the form $\epsilon ({\bf k},n)=kn+\epsilon _{0}({\bf k})$,
where the quantity $\epsilon _{0}({\bf k})$ gives the contribution of vacuum
fluctuations of the field).
In the vacuum state $\mid 0>$ (which is determined now by $C({\bf k},n)\mid
0>=0$)
field oscillators (and all field observables) are absent. The operator of total
energy
of the field may be generalized in a natural way as
\begin{equation}
E=\sum \epsilon ({\bf k},n)C^{+}({\bf k},n)C({\bf k},n).
\end{equation}

The connection with the standard field variables may be determined with the
help of operators which increase (decrease) the energy of system on $k$
$([E,A^{(+)}({\bf k})]_{-}= \pm
k A^{(+)}({\bf k}))$
\begin{equation}
A^{+}({\bf k})=\sum^{\infty }_{n=0}(n+1)^{1/2}C^{+}({\bf k},n+1)C({\bf k},n),
\end{equation}
\begin{equation}
A({\bf k})=\sum^{\infty }_{n=0}(n+1)^{1/2}C^{+}({\bf k},n)C({\bf k},n+1).
\end{equation}

It may be seen from (3.4)-(3.6) that the operators $A$ and $A^{+}$satisfy
the commutation relations
\begin{equation}
[A({\bf k}),A^{+}({\bf k}')]_{-}=N({\bf k})\delta ^{3}({\bf k}-{\bf k}'),
\end{equation}
where $N({\bf k})=\sum^{\infty }_{n=0}C^{+}({\bf k},n)C({\bf k},n)$ is the
complete number of spatial domains related to the wave number ${\bf k}$.
If one restricts oneself by the states (2.6) with $N({\bf k})=1$, then the
operators $A^{+}({\bf k})$ and $A({\bf k})$ will be surely coincided with the
standard operators of the creation and annihilation of scalar particles.

As it was mentioned in sec.2, the quantities $N({\bf k},n)=C^{+}({\bf
k},n)C({\bf k},n)$
must be determined by dynamics. However, they may be estimated under the
simple considerations. It is clear that in the absence of the gravitational
interaction the quantities $N({\bf k},n)$ remain constant. Then, for instance,
under the assumption of the bounded  density $N<\infty $ of oscillators
satisfying the Fermi statistics it is easy to find
that the occupation numbers corresponding to the ground state are
\begin{equation}N({\bf k},n)=\theta (\mu -\epsilon ({\bf k},n)),
\end{equation}
where $\theta (x)=\left\{\begin{array}{l}
0\hbox{ for }x<0\hbox{ and {\it 1} for }x>0\end{array}\right\} $, and
$\mu $ is determined via the full number of oscillators $N=\sum N({\bf k},n)$.
Using (3.8) one can found the number of oscillators corresponding
to the wave vector ${\bf k}$
\begin{equation}
N({\bf k})=\sum^{\infty }_{n=0}\theta (\mu -\epsilon ({\bf k},n))=[1+(\mu
-\epsilon _{0}({\bf k}))/k],
\end{equation}
here $[x]$ denotes the entire part of the number $x$. In particular, from (3.9)
one can see that $N({\bf k})=0$ as $\mu <\epsilon _{0}({\bf k})$.

\section{Concluding Remarks}

For the excited states formed by the acting of the operators $A^{+}({\bf k})$
on the ground state (3.8) the operator $N({\bf k})$ is the usual function
(3.9).
Let us consider the excitations of the field (scalar particles) described by
the thermal equilibrium state corresponding to the temperature $T$ (one could
expect that the spatial domains created near a singularity have the thermal
spectrum [8]). Then the correlation function for the potentials of the field
(3.1) takes form
\begin{equation}
<\varphi (x)\varphi (x+{\bf r})>=(2\pi ^{2})^{-1}\int \Phi ^{2}(k){\hbox{{\it
sinkr}}\over kr} {dk\over k},
\end{equation}
where $\Phi ^{2}(k)=k^{2}N(k){1\over 2}${\it cth}$({k\over 2T})$. In the
region of the wave numbers $k\ll (T,\mu )$ the spectrum of
fluctuations of the field is occurred to be the scale-independent:
$\Phi ^{2}(k)\approx ${\it TkN}$(k)=T\mu $.

We also notice that the ground state determined by the occupation numbers
(3.8) has a bounded energy density of the field which may be considered as a
"dark
matter". Besides, we note that the mentioned property of spectrum to be
scale-invariant on
large scales for the thermal equilibrium state, actually, does not depend on
the
statistics of the oscillators (i.e. upon the choosing of the sign $\pm $ in
(3.3), (2.5)).

\begin{center}{\bf Acknowledgments}\end{center}
I am grateful to A.A.Starobinsky, V.D.Ivashchuk and D.V.Turaev for useful
discussions and critics.

\end{document}